\documentclass[prd,aps,preprint,tightenlines,superscriptaddress]{revtex4}
\usepackage{epsfig}
\preprint{DOE/ER/40762-317} \preprint{UM-PP\#05-001}

\begin{document}

\title{Drell-Hearn-Gerasimov Sum-Rule for the Deuteron \\
in Nuclear Effective Field Theory}
\author{Jiunn-Wei Chen}
\email{jwc@phys.ntu.edu.tw}
\affiliation{Department of Physics and National Center for Theoretical Sciences at
Taipei, National Taiwan University, Taipei, Taiwan 10617}
\author{Xiangdong Ji}
\email{xji@physics.umd.edu}
\affiliation{Department of Physics, University of Maryland, College Park, Maryland 20742}
\author{Yingchuan Li}
\email{yli@physics.umd.edu}
\affiliation{Department of Physics, University of Maryland, College Park, Maryland 20742}
\date{\today }

\begin{abstract}
The Drell-Hearn-Gerasimov sum rule for the deuteron is studied in
nuclear effective field theory. The low-energy theorem for the
spin-dependent Compton amplitude $f_1(\omega)$ is derived to the
next-to-leading order in low-energy expansion. The spin-dependent
photodisintegration cross section $\sigma^P-\sigma^A$ is
calculated to the same order, and its contribution to the
dispersive integral is evaluated.
\end{abstract}

\maketitle

\vspace{0.5in}

The Drell-Hearn-Gerasimov (DHG) sum rule is a dispersive sum rule
which relates the anomalous magnetic moment of a system,
elementary or composite, to an integral over the spin-dependent
photo-production cross section $\sigma^P-\sigma^A$ \cite{dhg}. The
sum rule is derived from the low-energy theorem for the
spin-dependent forward Compton amplitude $f_1(\omega)$\cite{low}
and a dispersion relation. In recent years, because of rapid
technological advances, it becomes possible to study this sum rule
experimentally. For example, the experiments recently done at
Mainz, Bonn, and Jefferson Lab were motivated by checking this sum
rule for the proton and neutron \cite{exp}.

In this paper, we examine the DHG sum rule for the deuteron
(spin-1) in light of the nuclear effective field theory (EFT). The
DHG sum rule for the deuteron reads,
\begin{equation}
\frac{\pi^2\alpha_{\mathrm{em}}\kappa^2_D}{M_D^2} = \int^\infty_{\omega_{%
\mathrm{th}}} d\omega \frac{\sigma^P(\omega)-\sigma^A(\omega)}{\omega} \ ,
\label{ddhg}
\end{equation}
where $\kappa_D = 2\mu_D M_Dc/e\hbar - 2$ is the deuteron's anomalous
magnetic moment in unit of $e\hbar/(2M_Dc)$, and $M_D$ is the mass. Because
the deuteron's magnetic moment is $\mu_D = 0.857\mu_N$, where $%
\mu_N=e\hbar/2M_Nc$ is the nuclear magneton and $M_N$ the nucleon mass, $%
\kappa_D = -2\times 0.143$. Numerically, the left-hand side is
0.65$\mu$b. The $\omega_{\mathrm{th}}$ is the threshold photon
energy for the deuteron photodisintegration; and $\sigma^P$ and
$\sigma^A$ are the photo-production cross sections with the
helicity of the photon parallel or anti-parallel to the helicity
(+1) of the deuteron, respectively.

It has been realized for sometime that nuclear physics at low
energy might be understood by effective field theories (EFT) which
work according to the same principles as the standard model
\cite{weinberg}. However, constructing a workable scheme for
specific systems is not necessarily straightforward. In the past
few years, considerable progress has been made in two nucleon
sector (see \cite{review} for a recent review). It began with the
pioneering work of Weinberg, who proposed to encode the short
distance physics in the derivative expansion of local operators
\cite{weinberg}. The problem associated with the unusually small
binding energy of the deuteron was solved by Kaplan, Savage and
Wise by exploiting the freedom of choosing a renormalization
substraction scheme \cite{kaplan1}, quickly followed by the
pionless version \cite{chen1} (see also
\cite{vK97,Cohen97,BHvK1}). Requiring reproducing the residue of
the deuteron pole at next-to-leading order (NLO), a version with
accelerated convergence was suggested in \cite{phillips}. The use
of dibaryon fields as the auxiliary fields, at first introduced in
\cite{kaplan2}, was taken seriously in \cite{savage} which
simplified the calculation significantly.

Using the latest formulation of EFT for the two-nucleon system, we
study both the left and right hand sides of the DHG sum rule for
the deuteron. The low-energy theorem is verified to NLO in low
energy expansion. The spin-orbit interactions arising from
non-relativistic reduction turn out to play a significant role.
Then the spin-dependent photo-disintegration cross section is
computed to the same order. Although the leading-order result
depends only on the nucleon scattering parameters, the NLO depends
on two electromagnetic counter terms whose coefficients can be
determined by the magnetic moment of the deuteron and the rate for $%
n+p$ radiative capture. Finally, the photodisintegration
contribution to the DHG integral is evaluated.

The structure of the forward Compton scattering amplitude for a
general spin target is,
\begin{equation}
f=f_{0}\hat{\epsilon}^{\prime \ast }\cdot \hat{\epsilon}+f_{1}i\hat{\epsilon}%
^{\prime \ast }\times \hat{\epsilon}\cdot \vec{S}+f_{2}(\hat{k}\otimes \hat{k%
})^{(2)}\cdot (\vec{S}\otimes \vec{S})^{(2)}\hat{\epsilon}^{\prime \ast
}\cdot \hat{\epsilon}+...\
\end{equation}%
where $\hat{\epsilon}$($\hat{\epsilon}^{\prime }$) is the initial
(final) photon polarization, $\vec{S}$ is the angular momentum
operator of the
target, and $\otimes $ indicates a tensor coupling. The vector amplitude $%
f_{1}$ is related to those with the target magnetic quantum number $m_{S}$,
\begin{equation}
f_{1}=-\frac{3}{S(S+1)}\frac{1}{2S+1}\sum_{m_{S}}m_{S}f^{(m_{S})}\ .
\end{equation}%
The amplitude has a low-energy expansion \cite{low},
\begin{equation}
f_{1}=-\frac{\alpha _{\mathrm{em}}\kappa ^{2}}{4S^{2}M^{2}}\omega
+2\gamma \omega ^{3}+... \ ,
\end{equation}%
where the first term corresponds to the famous low-energy theorem with the
anomalous magnetic moment $\kappa $ defined as $\mu -2S$ \cite{low}, where $%
\mu $ is the magnetic moment in unit of $e\hbar /2Mc$. The next
term defines the forward spin-polarizability $\gamma$ which has
been studied in EFT in \cite{liji}.

In a pionless effective field theory for the deuteron \cite%
{BHvK1,kaplan2,savage}, the nucleon field $N$ and the
$^{3}S_{1}$-channel dibaryon field $t_{j}$ are introduced. The
leading-order effective lagrangian is
\begin{eqnarray}
\mathcal{L} &=&N^{\dagger }\left( iD_{0}+\frac{\mathbf{D}^{2}}{2M_{N}}%
\right) N-t_{j}^{\dagger }\left[ iD_{0}+\frac{\mathbf{D}^{2}}{4M_{N}}-\Delta %
\right] t_{j}  \nonumber \\
&&-y\left[ t_{j}^{\dagger }N^{T}P_{j}N+\mathrm{h.c.}\right] \ ,
\end{eqnarray}%
where $P_{i}=\tau _{2}\sigma _{2}\sigma _{i}/\sqrt{8}$ is the
$^{3}S_{1}$ two-nucleon projection operator  and $y$ is a coupling
constant between the
dibaryon and two-nucleon in the same channel. The covariant derivative is $%
\mathbf{D}=\mathbf{\partial}+ieQ\mathbf{A}$ with $Q=(1+\tau
^{3})/2$ as the charge operator and $\mathbf{A}$ the photon vector
potential. The NN scattering amplitude is reproduced by the
following choice of parameters
\begin{equation}
y^{2}=\frac{8\pi }{M_{N}^{2}r^{(^{3}S_{1})}} \ ,~~~~\Delta =\frac{2}{%
M_{N}r^{(^{3}S_{1})}}\left( \frac{1}{a^{(^{3}S_{1})}}-\mu \right)
\ ,
\end{equation}%
where $a^{(^{3}S_{1})}$ is the scattering length,
$r^{(^{3}S_{1})}$ is the effective range, and $\mu $ is the
renormalization scale. Similarly, one can introduce the dibaryon
field to describe the scattering in the $^{1}S_{0}$ channel as
well.
%$a^{(^1S_0)}=-23.714$ fm is the scattering length in the
%two-nucleon singlet $S_0$ channel and

\begin{figure}[t]
\epsfig{file=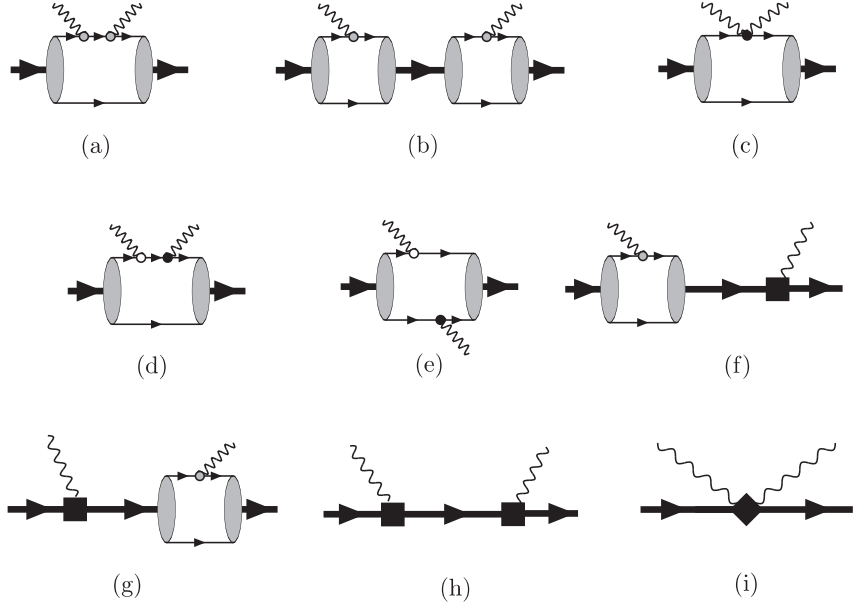,clip=,height=8.0cm,angle=0} \caption{Feynman
diagrams for spin-dependent forward Compton scattering on the
deuteron. The thick initial and final state arrows represent the
deuteron. The thick arrows in the middle of the diagrams denote
the dibaryon states in the $^3S_1$ and $^1S_0$ channels. The
shaded circles denote the magnetic moment interactions, and the
open circles the electric current interactions. The solid squares
are from electromagnetic counter terms ($L_1$ and $L_2$) at NLO.
The seagull vertices and the solid circles are from relativistic
spin-orbit interactions.}
\end{figure}

Let us compute the spin-dependent forward Compton amplitude $%
f_{1}(\omega )$ on the deuteron. The Feynman diagrams to NLO are
shown in Fig. 1, where the crossing diagrams are omitted. The
shaded circles represent the photon magnetic coupling with the
nucleon,
\begin{equation}
\mathcal{L}_{\mathrm{em}}^{\mathrm{LO}}=\frac{e}{2M_{N}}N^{\dagger }\left(
\mu ^{(0)}+\mu ^{(1)}\tau _{3}\right) \mathbf{\sigma }\cdot \mathbf{B}N
\end{equation}%
where $\mu ^{(0)}=(\mu _{p}+\mu _{n})/2$ and $\mu ^{(1)}=(\mu
_{p}-\mu _{n})/2$ are the isoscalar and isovector nucleon magnetic
moments in unit of nuclear magneton, $\mathbf{B}$ is an external
magnetic field. The contribution from the pure magnetic photon
coupling are shown in diagrams (a) and (b). A straightforward
calculation yields
\begin{equation}
f_{1}(\omega )|^{\mathrm{(a)+(b)}}=-\frac{e^{2}}{4\pi
M_{N}^{2}}(\mu ^{(0)})^{2}\omega +...\ ,
\end{equation}%
Although it appears as a N$^2$LO contribution in EFT power
counting, it is actually a leading-order one, proportional to
$q^2/\omega$, before setting the photon momentum $q=\omega$.

The magnetic coupling is generated from a relativistic
interaction, which, after non-relativistic reduction, also
produces a \textquotedblleft spin-orbit" interaction
\begin{equation}
\mathcal{L}_{\mathrm{em}}^{\mathrm{N^2LO,SO}}=N^{\dagger }i\left[
\left( 2\mu
^{(0)}-\frac{1}{2}\right) +\left( 2\mu ^{(1)}-\frac{1}{2}\right) \tau _{3}%
\right] \frac{e}{8M_{N}^{2}}\mathbf{\sigma }\cdot \left(
\mathbf{D}\times \mathbf{E}-\mathbf{E}\times \mathbf{D}\right) N\
,
\end{equation}
where $\mathbf{E}$ is an external electric field. The above
lagrangian contains a seagull interaction for the proton as shown
in diagram (c), which contributes to $f_1(\omega)$ at the same
order as (a) and (b) do. Moreover, there is a derivative coupling
which is shown as solid circles in diagrams in Fig. 1. This
coupling, when combined with a current interaction from the gauged
part of the proton's kinetic energy, generates a contribution to
$f_1(\omega)$ as shown in diagrams (d) and (e). The combined
result from the spin-orbit interaction acting on the proton is
\begin{equation}
f_{1}(\omega )|^{(c)+(d)}=\frac{e^{2}}{16\pi M_{N}^{2}}(2\mu
_{p}-1)\omega +... \ .
\end{equation}%
When the spin-orbit term acts on the neutron (e), the result is
proportional to the magnetic moment of the neutron,
\begin{equation}
f_{1}(\omega )|^{(e)}=\frac{e^{2}}{16\pi M_{N}^{2}}2\mu _{n}\omega
+... \ .
\end{equation}%
Summing over the above contributions, one has in EFT
\begin{equation}
f_{1}(\omega )|^{\mathrm{LO}}=-\frac{e^{2}}{16\pi M_{N}^{2}}(\mu
_{n}+\mu _{p}-1)^{2}\omega +... \ .
\end{equation}%
At this order, the magnetic moment and mass of the deuteron are
the sum of those of the neutron's and proton's, $\mu _{d}=\mu
_{n}+\mu _{p}$ nuclear magneton, $M_{D}=M_{p}+M_{n}\approx
2M_{N}$, respectively. The above result is clearly the same as the
low-energy theorem.

There is a new electromagnetic counter-term at NLO,
\begin{equation}
\mathcal{L}_{\mathrm{em,2}}^{\mathrm{NLO}} =
-i\frac{e}{M_{N}}\left( \mu^{(0)}
-\frac{L_{2}}{r^{(^{3}S_{1})}}\right) \varepsilon
^{ijk}t_{i}^{\dagger }B_{j}t_{k} \ ,
\end{equation}
where we introduce a $\mu^{(0)}$ term in the definition, chosen to
cancel the wave function renormalization contribution, which is
present in the leading order result in the di-baryon formulation
but has been omitted so far. The above contributes to the magnetic
moment of the deuteron is
\begin{equation}
\mu_d = 2\mu^{(0)} + \frac{2\gamma L_2}{1-\gamma r^{(^3S_1)}} \ ,
\label{dmm}
\end{equation}
in unit of nuclear magneton, where $\gamma=\sqrt{M_NB}=45.703$ MeV with $%
B=2.225$ MeV the deuteron binding energy. Fitting to the experimental value,
one finds, $L_2 = -0.03$ fm.

Using a solid square to denote the above interaction, its
contribution to the NLO Compton amplitude is shown by the three
Feynman diagrams (f-h) in Fig. 1. In addition, there is an
associated term from relativistic correction,
\begin{equation}
\mathcal{L}_{\mathrm{em,2}}^{\mathrm{N^2LO,SO}}=\frac{e}{2M_{N}^{2}}\left(
\mu ^{(0)}-\frac{L_{2}}{r^{(^{3}S_{1})}}-\frac{1}{4}\right)
\varepsilon ^{ijk}t_{i}^{\dagger }\left( \mathbf{D}\times
\mathbf{E}-\mathbf{E}\times \mathbf{D}\right) _{j}t_{k}\ ,
\end{equation}%
which generates a sea-gull contribution shown in diagram (i).
Summing over the above contributions, we find the spin-dependent
Compton amplitude to NLO,
\begin{equation}
f_{1}(\omega )|^{(\mathrm{LO+NLO)}}=-\frac{e^{2}}{4\pi (2M_{N})^{2}}(\mu
_{d}-1)^{2}\omega +...\ ,
\end{equation}%
where $\mu _{d}$ is the NLO result shown in eq. (\ref{dmm}). The
result is again consistent with the low-energy theorem. Additional
relativistic corrections will systematically converts $2M_{N}$
into a deuteron mass.

We now turn to the right-hand side of the DHG sum rule in eq.
(\ref{ddhg})---the photon-energy integration over the entire
spin-dependent production cross section from the threshold to
infinity. Experimentally, there have been preliminary
data from Mainz on meson production and more data will be analyzed soon \cite%
{mainzd}. But there is no direct data on the cross-section
asymmetry $\sigma^P-\sigma^A$ in the region of deuteron
photodisintegration. The HIGS facility at Duke University is
poised to make this measurement in the near future \cite{weller}.

There are theoretical estimates on the cross section asymmetry from nuclear
and hadronic models. A most complete and up-to-date study was made by Arenh%
\"{o}vel \cite{arenhovel1,arenhovel3}, who classifies the cross section into
three types:

\begin{itemize}
\item {photo-disintegration $\gamma+d\rightarrow n+p$,}

\item {single-pion production including coherent pion production $%
\gamma+d\rightarrow d+\pi^0$, and incoherent production $\gamma+d\rightarrow
N+N+\pi$,}

\item {two and more pion and other meson production.}
\end{itemize}

The estimate shows that the first process contributes about $-383$
$\mu$b to the DHG integral, the second 299 $\mu$b, and the third
70 $\mu$b. A large cancellation occurs the between meson
production and photodisintegration, as is dictated by the sum
rule. A simplified way of understanding the cancellation is to
imagine a complete scale separation between the deuteron structure
physics and the nucleon physics. Photoproduction, independently on
the proton and neut ron in the deuteron, yields a contribution
about 438 $\mu$b. This must be largely cancelled by the
photodisintegration contribution. In reality, of course, the
deuteron structure physics can strongly affect the outcome of
meson production in individual channels, even at very high energy.
For instance, a significant effect is coherence pion production,
which contributes about 99 $\mu$b to the integral according to
Ref. \cite{arenhovel3}. Moreover, charged single-pion production
off the proton and neutron is strongly modified by the final-state
interactions. It is likely, however, that the complete integral is
less sensitive to the deuteron structure and final-state
interaction effects.

The nuclear EFT allows calculation of the deuteron
photodisintegration cross section at low energy. Using the optical
theorem, one can obtain the cross section through the imaginary
part of the forward Compton scattering amplitude, for which
Feynman diagrams are again those shown in Fig. 1. At NLO, there is
an additional electromagnetic counter-term that couples the
$^3S_1$ and $^1S_0$ channels,
\begin{equation}
\mathcal{L}_{\mathrm{em, 1}}^{\mathrm{NLO}} = e\frac{L_{1}}{M_{N}\sqrt{%
r^{(^{1}S_{0})}r^{(^{3}S_{1})}}}t_{j}^{\dagger }s_{3}B_{j}+\mathrm{h.c.}
\end{equation}
where $s_a$ is the dibayron field with quantum number of isovector
$^1S_0$. The coupling constant $L_1$ has been determined by the rate of $%
n+p\rightarrow d+\gamma$. The measured cross section $\sigma =
334.2\pm 0.5$ mb with an incident neutron speed of $2200$ m/s
fixes $L_1=-4.42$ fm.

\begin{figure}[t]
\epsfig{file=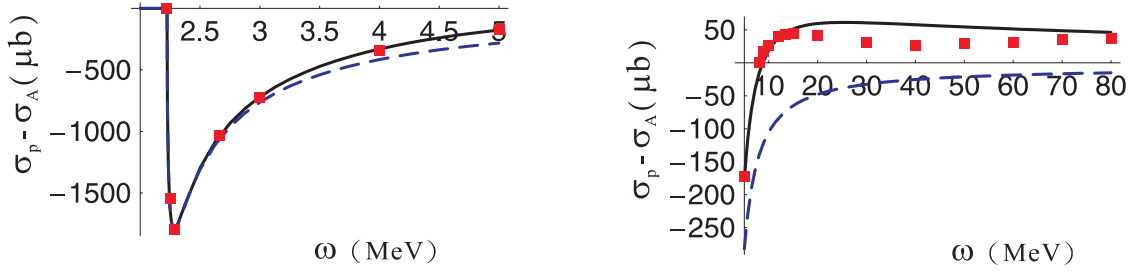,clip=,height=3.5cm,angle=0} \caption{Cross
section asymmetry for the deuteron photodisintegration, shown in
two difference photon-energy scale. The solid line is the full EFT
result to NLO, the dashed line does not contain the relativistic
spin-orbit effects. The squares show the potential model result of
Arenhovel \cite{arenhovel1}. }
\end{figure}

To NLO, the cross section difference for the photo-disintegration is
\begin{eqnarray}
\left. \sigma^{P}-\sigma^{A}\right\vert_{\vec{\gamma } \vec{d}\rightarrow
np} &=&-\frac{e^{2}p\gamma }{2M_{N}^{2}\left( p^{2}+\gamma ^{2}\right)
\left( 1-\gamma r^{(^{3}S_{1})}\right) }  \nonumber \\
&&\times \left\{ \frac{\left[ 2\mu^{(1)}\left( \gamma -1/a^{(^{1}S_{0})}+%
\frac{1}{2}r^{(^{1}S_{0})}p^{2}\right) +\left( p^{2}+\gamma ^{2}\right) L_{1}%
\right] ^{2}}{\left( -1/a^{(^{1}S_{0})}+\frac{1}{2}r^{(^{1}S_{0})}p^{2}%
\right) ^{2}+p^{2}}\right.  \nonumber \\
&&\left.+\frac{\left( p^{2}+\gamma ^{2}\right) L_{2}^{2}}{1-\gamma
r^{(^{3}S_{1})}+r^{(^{3}S_{1})^{2}}\left( p^{2}+\gamma ^{2}\right) /4}
-4\left( 4\mu^{(1)}-1\right) \frac{p^{2}}{3\left( p^{2}+\gamma ^{2}\right) }
\right\} \ ,
\end{eqnarray}%
where $p=\sqrt{M_{N}\omega -\gamma ^{2}}$ and $\omega $ is the
photon energy. The first term in the braces comes from the
production of the $^1S_0$ proton-neutron scattering state, and the
second term from the $^3S_1$ scattering state. In the latter case,
the contribution from the magnetic coupling alone vanishes because
of the orthogonality of the scattering and bound state wave
functions. The third term comes from the interference between the
current interaction and the spin-orbit term, involving multiple
nucleon-nucleon partial waves in the final state.

In Fig. 2, we have shown, using solid lines, the cross section
difference as a function of photon energy in two different scales.
Near the threshold, the dominant contribution comes from the
$^{1}S_{0}$ final state. It is negative because only an
anti-parallel photon-deuteron configuration has a non-zero cross
section. The $L_{2}$ contribution is small throughout the region.
When the photon energy is 10 MeV and higher, the spin-orbit
interaction becomes significant. It fact, it changes the cross
section asymmetry from positive to negative, as is made clear by
the difference between the solid and dashed lines. The potential
model calculation in Ref. \cite{arenhovel3} is shown by the
squares, for the sake of clarity. The agreement between the EFT
and model calculation is excellent blow $\omega = 10$ MeV. The
former becomes less trustworthy at higher photon energy.

Finally, we consider the contribution of photodisintegration to
the DHG integral in EFT. In principle, one should cut-off the
integral at some photon-energy, beyond which the nuclear EFT is no
longer applicable. Indeed the model calculation shows a strong
effect from the $\Delta$ resonance \cite%
{arenhovel3}, which is clearly beyond the nuclear EFT. However,
because the integral is manifestly convergent, we extend the
integration all the way to infinity as our crude estimate. This
yields a contribution $-385$ $\mu$b to the DHG integral.

To summarize, we verify that the nuclear EFT reproduces the
low-energy theorem for the spin-dependent deuteron Compton
amplitude. In the same framework, we calculate the spin-dependent
photo-disintegration cross section, which has been compared to a
potential model calculation and can be tested by future
experimental data.

This work was supported by the U. S. Department of Energy via grant
DE-FG02-93ER-40762 and by the National Science Council of ROC.

\end{document}